\documentclass[french,english]{revtex4-1}
\usepackage[T1]{fontenc}
\usepackage[latin9]{inputenc}
\usepackage{textcomp}
\usepackage{amsthm}
\usepackage{amsmath}
\usepackage{amssymb}
\usepackage{esint}

\makeatletter
\numberwithin{equation}{section}
\numberwithin{figure}{section}

\usepackage[colorlinks = true,
            linkcolor = red,
            urlcolor  = blue,
            citecolor = blue,
            anchorcolor = blue]{hyperref}
\makeatother

\usepackage{babel}
\makeatletter
\addto\extrasfrench{%
   \providecommand{\fg}{\ifdim\lastskip>\z@\unskip\fi~\frqq}%
}

\makeatother
\begin{document}

\title{ANALYTICAL SOLUTION OF (2+1) DIMENSIONAL DIRAC EQUATION IN TIME-DEPENDENT
NONCOMMUTATIVE PHASE-SPACE }

\author{{\normalsize{}Ilyas Haouam}}

\email{ilyashaouam@live.fr}

\address{Laboratoire de Physique Mathématique et de Physique Subatomique (LPMPS),
Université Frères Mentouri, Constantine 25000, Algeria}
\begin{abstract}
{\normalsize{}In this article, we studied the system of (2+1) dimensional
Dirac equation in time-dependent noncommutative phase-space. Exactly,
we investigated the analytical solution of the corresponding system
by the Lewis-Riesenfeld invariant method based on the construction
of the Lewis-Riesenfeld invariant. Knowing that we obtained the time-dependent
Dirac Hamiltonian of the problem in question from a time-dependent
Bopp-Shift translation, then it used to set the Lewis-Riesenfeld invariant
operators. Thereafter, the obtained results used to express the eigenfunctions
that lead to determining the general solution of the system.}{\normalsize \par}

$\phantom{}$

$\phantom{}$

$\phantom{}$

\end{abstract}


\maketitle

\section{{\normalsize{}Introduction}}

It is known that Heisenberg suggested the opinion of noncommutative
(NC) space-time in 1930, and in 1947, Snyder presented it \cite{key-1,key-2}
to the necessity to regularizing the divergence of the quantum field
theory. Then, in recent years, noncommutative geometry (NCG) became
very interesting for studying several physical problems, and it became
clear that there is a strong connection between NCG and string theories.
Studies of this geometric type and its involvement have been incorporated
with important physical concepts and tools, and have been useful in
highlighting various fields of physics, particularly in matrix theory
(matrix model BFSS (1997)) \cite{key-3}. NCG involved also in the
description of quantum gravity theories \cite{key-4}, Aharonov-Bohm
effect \cite{key-5}, Aharonov-Casher effect \cite{key-6}, etc \cite{key-7}.
Knowing that the origins of NCG related to the investigations for
topological spaces of C\textasteriskcentered -algebras of functions.
Later this type of geometry was theorized by A. Connes and others
in 1985 \cite{key-8,key-9,key-10,key-11,key-12} by studying and defining
a cyclical cohomology. It has been shown that the differential calculus
on manifolds had a NC equivalent. Next, the NCG found great encouragement
through several mathematical results such as of K-theory of \foreignlanguage{french}{C\textasteriskcentered -}algebras,
\foreignlanguage{french}{Gelfand-Naïmark} theorem on the \foreignlanguage{french}{C\textasteriskcentered -}algebras,
characterizations of commutative von Neumann algebras, cyclic cohomology
of the \foreignlanguage{french}{$C^{\infty}(M)$} algebra, relations
between Dirac operators and Riemannian metrics, Serre-Swan theorem,
etc. The idea of phase-space noncommutativity is largely motivated
by the foundations of quantum mechanics through the canonical quantization.

It is easy to apply the phase-space noncommutativity using the ordinary
product with Weyl operators (Weyl-Wigner maps)\cite{key-12}, or by
replacing the ordinary product with the Moyal-Weyl product ($\star$-product)
in the functions and actions of our systems\cite{key-13,key-14},
also the Bopp-shift linear transformations \cite{key-15,key-16},
and the Seiberg-Witten maps \cite{key-8,key-10,key-17}. 

Studying physics within NCG has attracted a lot of interest in recent
years, because noncommutativity is necessary when considering the
low-energy efficiency of D-brane with a background magnetic field,
and also in a tiny scale of strings or in conditions of the very high
energy, the effects of noncommutativity may appear. Besides, one of
the strong motivations of NCG, is to obtain a coherent mathematical
framework in which it would be possible to describe quantum gravitation.
For all these reasons and advantages, we carry out this work in NC
formalism. In addition, it is interesting to find other models in
which noncommutativity emerges.

Several scientific works have focused on the time-independent noncommutativity.
Experimental research has considered the parameters of the noncommutativity
of fixed values in the context of cosmic microwave background radiation,
perhaps considered approximately fixed to the celestial sphere, for
example, as proposed in ref \cite{key-18}. However,  differently,
in our work, our obvious intention is to involve the time-dependency
in NC parameters because of the possibility that NC parameters may
show time-dependency. For instance, physical measurements must take
into account the effect of the Earth\textquoteright s rotation around
its axis, which causes a time-dependency in the NC parameters.

On the other hand, the motivations for choosing to study the (2+1)
dimensional Dirac equation are due to several important works in this
context, such as the investigations of Landau levels \cite{key-19},
the oscillation of magnetization \cite{key-20}, Weiss oscillation
\cite{key-21}, de Haas-van Alphen effect \cite{key-22}, analysis
through coherent states \cite{key-23}, the movement of electrons
transporters in graphene and other materials \cite{key-24}, etc.
Particularly, the 2 dimensional Dirac equation in interaction with
a homogeneous magnetic field has various applications in graphene
such as in refs \cite{key-25,key-26}, and in studying quantum Hall
effect and fractional Hall effects \cite{key-27,key-28}, Berry phase
\cite{key-29}, etc. In graphene and other materials such as in Weyl
semimetals, an important phenomenon takes place if the magnetic field
and the uniform electric field are introduced. Exactly, the spacing
between different Landau levels decreases if the electric field strength
reaches a critical value \cite{key-30,key-31}.

In our study about (2+1) dimensional Dirac system, the noncommutativity
will be considered time-dependent through a time-dependent Darboux
transformation ``Bopp's shift''. This, in turn, makes the studied
system a time-dependent one, $\mathcal{H}(x_{i}^{nc},p_{i}^{nc})\longrightarrow\mathcal{H}(t)$.

Solving the system of equations in interaction with time-dependent
potentials has attracted many physicists for a long time. In addition
to the essential mathematical benefit, this topic is related to a
lot of physical problems and applications for its applicability. For
instance in quantum transport \cite{key-32,key-33,key-34}, quantum
optics \cite{key-35,key-36,key-37}, quantum information \cite{key-38},
the degenerate parametric amplifier \cite{key-39}, also spintronics
\cite{key-40,key-41}, and in the description of the two trapped cold
ions dynamics in the Paul trap \cite{key-42}. To study systems of
time-dependent equations, there are many methods like the evolution
operator, the change of representation, the unitary transformations,
path integral, second quantization, Lewis-Riesenfeld (LR) invariant
approach. Also, there are other used techniques, as in refs \cite{key-43,key-44}. 

The LR method \cite{key-45,key-46} is a technique that allows to
obtaining a set of solutions of time-dependent equation systems, through
the eigenstates of LR invariants. These invariants are built to find
the solutions of such systems of equations, where Lewis and Riesenfeld
in their original paper presented a technique to obtaining a group
of exact wave-functions for the time-dependent harmonic oscillator
in Hilbert space. The LR approach has been applied in many applications
such as in mesoscopic R(t)L(t)C(t) electric circuits where the quantum
evolution is described\cite{key-47}. As well in in engineering, in
shortcuts and adiabaticity \cite{key-48}...

A large variety of scientific papers concerning time-dependent systems
were interested in the time-dependent harmonic oscillator, or in time-dependent
linear potentials, but in our current work, to be more specific we
report the time-dependent Background of the NC phase-space. We consider
a time-dependent Bopp-shift translation to transform the system to
a time-dependent NC one, then due to the LR invariant method we obtain
the LR invariant and its eigenstates to solve our system equations.

\section{{\normalsize{}time-dependent noncommutativity }}

In the theory of NCG space may not commute anymore (i.e. AB\ensuremath{\neq}BA).
In a $d$ dimensional time-dependent NC phase-space let us consider
the operators of coordinates and momentum $x_{j}^{nc}$ and $p_{k}^{nc}$
respectively. These NC operators satisfy the deformed commutation
relations \cite{key-49}
\begin{equation}
\begin{array}{cccc}
\left[x_{j}^{nc},x_{k}^{nc}\right] & = & i\Theta_{jk}(t)\\
\left[p_{j}^{nc},p_{k}^{nc}\right] & = & i\eta_{jk}(t), & (j,k=1,..,d)\\
\left[x_{j}^{nc},p_{k}^{nc}\right] & = & i\hbar^{eff}\delta_{jk}
\end{array},\label{eq:1}
\end{equation}
the effective Planck constant being
\begin{equation}
\hbar^{eff}=\hbar\left(1+\frac{\Theta\eta}{4\hbar^{2}}\right),\label{eq:2}
\end{equation}
where $\frac{\Theta\eta}{4\hbar^{2}}\ll1$ is the consistency condition
in the usual quantum mechanics. $\delta_{jk}$ is the identity matrix,
and $\Theta_{jk}$, $\eta_{jk}$ are real constant antisymmetric $d\times d$
matrices.

In some studies concerning the NC parameters  as in the experiment
by\textit{ }``Nesvizhevsky et al'' \cite{key-50,key-51}, we note
that $\Theta\approx10^{-30}m^{2}$ and $\eta\approx1,76.10^{-61}Kg^{2}m^{2}s^{-2}$.
Other bounds exist. For example $\Theta\approx4.10^{-40}m^{2}$ when
assuming the natural units, $\hbar=c=1$ \cite{key-52}. As well as
when taking into account that the experimental energy resolution is
related to the uncertainty principle because of the finite lifetime
of the neutron, this leads to obtaining $\eta\approx10^{-67}Kg^{2}m^{2}s^{-2}$
(kind of a correction). These obtained results including of the experiment
by ``Nesvizhevsky et al'', allow us to evaluate the consistency
condition of the NC model $\left|\frac{\Theta\eta}{4\hbar^{2}}\right|\preceq10^{-24}$.
But if we consider the modifications introduced by noncommutativity
over $\hbar$ value (the precision is about $10^{-9}$), which are
at least about 24 orders of magnitude smaller than its value, with
considering the corrected bounds of $\eta$, we have $\left|\frac{\Theta\eta}{4\hbar^{2}}\right|\preceq10^{-29}$
\cite{key-53}. These values agree with the higher limits on the basic
scales of coordinate and momentum. These limits will be suppressed
if the used magnetic field in the experiment is weak, about $B\approx5mG$. 

As long as the system in which we investigate the effects of noncommutativity
is 2 dimensional, we restrict ourselves to the following NC algebra
\begin{equation}
\begin{array}{cccc}
\left[x_{j}^{nc},x_{k}^{nc}\right] & = & i\Theta e^{\gamma t}\epsilon_{jk}\\
\left[p_{j}^{nc},p_{k}^{nc}\right] & = & i\eta e^{-\gamma t}\epsilon_{jk},\; & (j,k=1,2)\\
\left[x_{j}^{nc},p_{k}^{nc}\right] & = & i\hbar^{eff}\delta_{jk}
\end{array},\label{eq:3}
\end{equation}
\foreignlanguage{french}{we have $\epsilon_{12}=-\epsilon_{21}=1$,
$\epsilon_{11}=\epsilon_{22}=0$, and $\Theta$, $\eta$ }are real-valued
with the dimension of $length{}^{2}$ and $momentum{}^{2}$, respectively.

While the space coordinates and momentum are fuzzy and fluid \cite{key-54},
they can not be localized, unless for minus infinite times. The parameters
$\Theta$, $\eta$ represent the fuzziness and $\gamma$ represents
the fluidity of the space. The above equation is the relation of the
ordinary NCG except that NC structure constants are considered as
exponentially increasing functions with the evolution of time. Certainly,
there are a multitude of other possibilities, such as $\Theta(t)=\Theta cos(\gamma t)$,
$\eta(t)=\eta sin(\gamma t)$. 

The new deformed geometry can be described by the operators
\begin{equation}
\begin{array}{cc}
x_{1}^{nc}=x^{nc}=x-\frac{1}{2\hbar}\Theta e^{\gamma t}p_{y},\: & p_{1}^{nc}=p_{x}^{nc}=p_{x}+\frac{1}{2\hbar}\eta e^{-\gamma t}y\\
x_{2}^{nc}=y^{nc}=y+\frac{1}{2\hbar}\Theta e^{\gamma t}p_{x},\: & p_{2}^{nc}=p_{y}^{nc}=p_{y}-\frac{1}{2\hbar}\eta e^{-\gamma t}x
\end{array}.\label{eq:5}
\end{equation}

When $\gamma=0$, the time-dependency in the structure of NC parameters
vanishes. In Addition, for $\Theta=\eta=0$, the NCG reduces to commutative
one, yonder the coordinates $x_{j}$ and the momentum $p_{k}$ satisfy
the ordinary canonical commutation relations
\begin{equation}
\begin{array}{cccc}
\left[x_{j},x_{k}\right] & = & 0\\
\left[p_{j},p_{k}\right] & = & 0, & (j,k=1,2)\\
\left[x_{j},p_{k}\right] & = & i\hbar\delta_{jk}
\end{array}.\label{eq:4}
\end{equation}

\section{{\normalsize{}(2+1) D Explicitly time-dependent Dirac equation and
its invariant operator}}

\subsection{(2+1) D Dirac equation in time-dependent noncommutative phase-space }

In presence of an electromagnetic four-potential $A_{\mu}=(A_{0},A_{i}),$
the Dirac equation in (2+1) d is given by

\selectlanguage{french}%
\begin{equation}
\begin{array}{c}
\left(c\alpha_{i}(p_{i}-\frac{e}{c}A_{i}(x))+eA_{0}(x)+\beta mc^{2}\right)\left|\psi\right\rangle =i\hbar\frac{\partial}{\partial t}\left|\psi\right\rangle ,\end{array}\label{eq:6}
\end{equation}
\foreignlanguage{english}{with $\left|\psi\right\rangle $ is the
Dirac wave function, and $p_{j}=\left(\begin{array}{c}
p_{x}\\
p_{y}\\
p_{z}
\end{array}\right)$ is the momentum. The Dirac matrices $\alpha_{j}$ , $\beta$
\begin{equation}
\alpha_{j}=\left(\begin{array}{cc}
0 & \sigma_{j}\\
\sigma_{j} & 0
\end{array}\right),\alpha_{1}=\sigma_{1}=\left(\begin{array}{cc}
0 & 1\\
1 & 0
\end{array}\right),\alpha_{2}=\sigma_{2}=\left(\begin{array}{cc}
0 & -i\\
i & 0
\end{array}\right),\beta=\sigma_{3}=\left(\begin{array}{cc}
1 & 0\\
0 & -1
\end{array}\right),\;\mathcal{I}=\left(\begin{array}{cc}
1 & 0\\
0 & 1
\end{array}\right),\label{eq:6-1}
\end{equation}
satisfy the following anticommutation relations
\begin{equation}
\left\{ \alpha_{i},\alpha_{j}\right\} =2\delta_{ij}\:,\;\left\{ \alpha_{i},\beta\right\} =0\mbox{ with }\alpha_{i}^{2}=\beta^{2}=1.\label{eq:7}
\end{equation}
}

\selectlanguage{english}%
We consider the magnetic field $\overrightarrow{B}$ along z-direction,
and it is defined in terms of the symmetric potential
\begin{equation}
A_{i}=\frac{B}{2}\left(-y,x,0\right),\mbox{ with }A_{0}=0,\label{eq:8}
\end{equation}
most research about time-dependent systems concerns the presence of
an electric field. Differently, in our current work, we do not rely
on the electric field. 

Using Eq.(\ref{eq:8}), then the Hamiltonian of the system becomes
\begin{equation}
\mathcal{H}\left(x,y,p_{x}p_{y}\right)=c\alpha_{1}p_{x}+c\alpha_{2}p_{y}+e\alpha_{1}\frac{B}{2}y-e\alpha_{2}\frac{B}{2}x+\beta mc^{2}.\label{eq:9}
\end{equation}

Achieving the NCG in the Dirac Hamiltonian (\ref{eq:9}) as follows

\begin{equation}
\mathcal{H}\left(x^{nc},y^{nc},p_{x}^{nc},p_{y}^{nc}\right)=c\alpha_{1}p_{x}^{nc}+c\alpha_{2}p_{y}^{nc}-e\alpha_{2}\frac{B}{2}x^{nc}+e\alpha_{1}\frac{B}{2}y^{nc}+\beta mc^{2},\label{eq:10}
\end{equation}
by applying Eq.(\ref{eq:5}), we necessarily express the new NC Hamiltonian
using the commutative variables $\left\{ x,y,p_{x},p_{y}\right\} $,
and by assuming that $\hbar=c=1$ (natural units) to simplify the
calculations, then we obtain
\begin{equation}
\mathcal{H}^{nc}(x,y,p_{x}p_{y},t)=\alpha_{1}(1+\frac{eB}{4}\Theta e^{\gamma t})p_{x}-\alpha_{2}(\frac{eB}{2}+\frac{\eta}{2}e^{-\gamma t})x+\alpha_{2}(1+\frac{eB}{4}\Theta e^{\gamma t})p_{y}+\alpha_{1}(\frac{eB}{2}+\frac{\eta}{2}e^{-\gamma t})y+\beta m.\label{eq:11}
\end{equation}

The time-dependent Dirac equation in NC phase-space is giving by
\begin{equation}
i\frac{\partial}{\partial t}\left|\bar{\psi}\left(t\right)\right\rangle =\mathcal{H}^{nc}\left(t\right)\left|\bar{\psi}\left(t\right)\right\rangle ,\label{eq:12}
\end{equation}
where $\left|\bar{\psi}\left(t\right)\right\rangle $ is the Dirac
NC wave function.

\subsection{The construction of the Lewis-Riesenfeld invariants }

To solve Eq.(\ref{eq:12}), we use the LR invariant method, which
assumes the existence of a quantum-mechanical invariant I(t) which
satisfies

\begin{equation}
\frac{dI(t)}{dt}=-i\left[I(t),\mathcal{H}^{nc}\left(t\right)\right]+\frac{\partial I(t)}{\partial t}=0,\label{eq:13}
\end{equation}
with
\begin{equation}
i\frac{\partial}{\partial t}\left(I(t)\left|\bar{\psi}\left(t\right)\right\rangle \right)=\mathcal{H}^{nc}\left(t\right)I(t)\left|\bar{\psi}\left(t\right)\right\rangle .\label{eq:13-1-1}
\end{equation}

The Eq.(\ref{eq:13}) is called the invariance condition for the dynamical
invariant operator I(t), which is a Hermitian operator
\begin{equation}
I(t)=I^{+}(t).\label{eq:13-1}
\end{equation}

Assuming that
\begin{equation}
I(t)=A_{1}(t)p_{x}+B_{1}(t)x+A_{2}(t)p_{y}+B_{2}(t)y+C(t),\label{eq:14}
\end{equation}
with \foreignlanguage{french}{$A_{1}(t)$, $B_{1}(t)$, $A_{2}(t)$,
$B_{2}(t)$, $C(t)$ }are time-dependent matrices. The substitution
of Eqs.(\ref{eq:14}, \ref{eq:10}) into Eq.(\ref{eq:13}), and using
the properties of the commutation relations, lead to
\begin{equation}
\left[I,\mathcal{H}^{nc}\right]+i\frac{\partial I}{\partial t}=\left[A_{1}p_{x},\mathcal{H}^{nc}\right]+\left[B_{1}x,\mathcal{H}^{nc}\right]+\left[A_{2}p_{y},\mathcal{H}^{nc}\right]+\left[B_{2}y,\mathcal{H}^{nc}\right]+\left[C,\mathcal{H}^{nc}\right]+i\frac{\partial I}{\partial t}=0,\label{eq:15}
\end{equation}
for simplicity we take $f_{\Theta}(t)=1+\frac{eB}{4}\Theta e^{\gamma t}$
and $f_{\eta}(t)=\frac{eB}{2}+\frac{\eta}{2}e^{-\gamma t}$, which
are not matrices, then we have
\begin{equation}
\begin{array}{c}
\left[A_{1},\alpha_{1}f_{\Theta}\right]p_{x}^{2}+\left[A_{2},\alpha_{2}f_{\Theta}\right]p_{y}^{2}-\left[B_{1},\alpha_{2}f_{\eta}\right]x^{2}+\left[B_{2},\alpha_{1}f_{\eta}\right]y^{2}+\left\{ \left[A_{1},\alpha_{2}f_{\Theta}\right]+\left[A_{2},\alpha_{1}f_{\Theta}\right]\right\} p_{x}p_{y}\\
+\left\{ \left[A_{1},\alpha_{1}f_{\eta}\right]+\left[B_{2},\alpha_{1}f_{\Theta}\right]\right\} yp_{x}+\left\{ \left[A_{1},\beta m\right]+\left[C,\alpha_{1}f_{\Theta}\right]+i\frac{\partial A_{1}}{\partial t}\right\} p_{x}+\left\{ \left[A_{2},\beta m\right]+\left[C,\alpha_{2}f_{\Theta}\right]+\frac{\partial A_{2}}{\partial t}\right\} p_{y}\\
+\left\{ \left[B_{1},\beta m\right]-\left[C,\alpha_{2}f_{\eta}\right]+i\frac{\partial B_{1}}{\partial t}\right\} x+\left\{ \left[B_{2},\beta m\right]+\left[C,\alpha_{1}f_{\eta}\right]+i\frac{\partial B_{2}}{\partial t}\right\} y+\left\{ -\left[A_{1},\alpha_{2}f_{\eta}\right]+\left[B_{1},\alpha_{1}f_{\Theta}\right]\right\} xp_{x}\\
+\left\{ \left[B_{1},\alpha_{2}f_{\Theta}\right]-\left[A_{2},\alpha_{2}f_{\eta}\right]\right\} xp_{y}+\left\{ \left[B_{1},\alpha_{1}f_{\eta}\right]-\left[B_{2},\alpha_{2}f_{\eta}\right]\right\} xy+\left\{ \left[A_{2},\alpha_{1}f_{\eta}\right]+\left[B_{2},\alpha_{2}f_{\Theta}\right]\right\} yp_{y}\\
+iA_{1}\alpha_{2}f_{\eta}+iB_{1}\alpha_{1}f_{\Theta}-iA_{2}\alpha_{1}f_{\eta}+iB_{2}\alpha_{2}f_{\Theta}-i\left[B_{1},\alpha_{1}f_{\Theta}\right]-i\left[B_{2},\alpha_{2}f_{\Theta}\right]+\left[C,\beta m\right]+i\frac{\partial C}{\partial t}=0
\end{array}.\label{eq:15-1}
\end{equation}

Then, to satisfy Eq.(\ref{eq:13}), and always taking advantage of
the properties of commutation relations, with $p_{i}p_{j}=p_{j}p_{i},\:x_{i}p_{j}=p_{j}x_{i}$
if i\ensuremath{\neq}j $\in$\{1,2\}, else $\begin{array}{c}
p_{x}x=xp_{x}-i,\:p_{y}y=yp_{y}-i\end{array}$. We demand 

\begin{equation}
\left[A_{1},\alpha_{1}f_{\Theta}\right]=0,\label{eq:25a}
\end{equation}
\begin{equation}
\left[A_{2},\alpha_{2}f_{\Theta}\right]=0,\label{eq:25b}
\end{equation}
\begin{equation}
\left[B_{1},\alpha_{2}f_{\eta}\right]=0,\label{eq:25c}
\end{equation}
\begin{equation}
\left[B_{2},\alpha_{1}f_{\eta}\right]=0,\label{eq:25d}
\end{equation}
\begin{equation}
\left[A_{1},\beta\right]m+\left[C,\alpha_{1}f_{\Theta}\right]+i\frac{\partial A_{1}}{\partial t}=0,\label{eq:25e}
\end{equation}
\begin{equation}
\left[A_{2},\beta\right]m+\left[C,\alpha_{2}f_{\Theta}\right]+i\frac{\partial A_{2}}{\partial t}=0,\label{eq:25f}
\end{equation}
\begin{equation}
\left[B_{1},\beta\right]m-\left[C,\alpha_{2}f_{\eta}\right]+i\frac{\partial B_{1}}{\partial t}=0,\label{eq:25g}
\end{equation}
\begin{equation}
\left[B_{2},\beta\right]m+\left[C,\alpha_{1}f_{\eta}\right]+i\frac{\partial B_{2}}{\partial t}=0,\label{eq:25h}
\end{equation}
\begin{equation}
\left[A_{1},\alpha_{2}f_{\Theta}\right]+\left[A_{2},\alpha_{1}f_{\Theta}\right]=0,\label{eq:25i}
\end{equation}
\begin{equation}
\left[B_{1},\alpha_{1}f_{\Theta}\right]-\left[A_{1},\alpha_{2}f_{\eta}\right]=0,\label{eq:25j}
\end{equation}
\begin{equation}
\left[B_{1},\alpha_{2}f_{\Theta}\right]-\left[A_{2},\alpha_{2}f_{\eta}\right]=0,\label{eq:25k}
\end{equation}
\begin{equation}
\left[B_{1},\alpha_{1}f_{\eta}\right]-\left[B_{2},\alpha_{2}f_{\eta}\right]=0,\label{eq:25l}
\end{equation}
\begin{equation}
\left[B_{2},\alpha_{1}f_{\Theta}\right]+\left[A_{1},\alpha_{1}f_{\eta}\right]=0,\label{eq:25m}
\end{equation}
\begin{equation}
\left[A_{2},\alpha_{1}f_{\eta}\right]+\left[B_{2},\alpha_{2}f_{\Theta}\right]=0,\label{eq:25n}
\end{equation}
\begin{equation}
iA_{1}\alpha_{2}f_{\eta}+iB_{1}\alpha_{1}f_{\Theta}-iA_{2}\alpha_{1}f_{\eta}+iB_{2}\alpha_{2}f_{\Theta}-i\left\{ \left[B_{1},\alpha_{1}f_{\Theta}\right]+\left[B_{2},\alpha_{2}f_{\Theta}\right]\right\} +\left[C,\beta m\right]+i\frac{\partial C}{\partial t}=0.\label{eq:25o}
\end{equation}

From the relations (\ref{eq:25a} - \ref{eq:25d}), and as long as
from Eq.(\ref{eq:25a}), we have
\begin{equation}
A_{1}=a_{0}(t)+a_{1}(t)\alpha_{1}+a_{2}(t)\alpha_{1}^{2}+a_{3}(t)\alpha_{1}^{3}+a_{4}(t)\alpha_{1}^{4}+...=a_{0}^{'}(t)+a_{1}^{'}(t)\alpha_{1},\mbox{ with }a_{i}^{'}(t)=a_{i}(t),\label{eq:26-111}
\end{equation}
therfore, we obtain

\begin{equation}
A_{1}=a_{1}+a_{2}\alpha_{1},\label{eq:40}
\end{equation}
\begin{equation}
A_{2}=a_{3}+a_{4}\alpha_{2},\label{eq:40b}
\end{equation}
\begin{equation}
B_{1}=b_{1}+b_{2}\alpha_{2},\label{eq:40c}
\end{equation}
\begin{equation}
B_{2}=b_{3}+b_{4}\alpha_{1}.\label{eq:40e}
\end{equation}

From Eqs.(\ref{eq:25e} - \ref{eq:25h}) and with the same manner,
supposing that $C$ is written in terms of $\alpha_{1}$, $\alpha_{2}$
and $\beta$ as follows
\begin{equation}
C=c_{1}+c_{2}\alpha_{1}+c_{3}\alpha_{2}+c_{4}\beta,\label{eq:40f}
\end{equation}
where $a_{j}$, $b_{j}$ and c$_{j}$ (with $j=1,..,4$) are supposed
to be time-dependent arbitrary functions. Substituting Eqs.(\ref{eq:40f},
\ref{eq:40}) into Eq.(\ref{eq:25e}) and Eqs.(\ref{eq:40f}, \ref{eq:40b})
into Eq.(\ref{eq:25f}), taking into consideration Eq.(\ref{eq:7})
yield
\begin{equation}
\begin{array}{ccc}
\frac{\partial a_{1}}{\partial t}=0,\: & \frac{\partial a_{3}}{\partial t}=0,\: & a_{2}=a_{4}=c_{2}=c_{3}=c_{4}=0\end{array},\label{eq:45}
\end{equation}
thereafter, substituting Eqs.(\ref{eq:40f}, \ref{eq:40c}) into Eq.(\ref{eq:25g})
and Eqs.(\ref{eq:40f}, \ref{eq:40e}) into Eq.(\ref{eq:25h}), taking
into consideration Eq.(\ref{eq:7}) yield 
\begin{equation}
\begin{array}{ccc}
\frac{\partial b_{1}}{\partial t}=0,\; & \frac{\partial b_{3}}{\partial t}=0,\: & b_{2}=b_{4}=c_{2}=c_{4}=0\end{array}.\label{eq:48}
\end{equation}

From the Eqs.(\ref{eq:45}, \ref{eq:48}) we note that $a_{1}$, $a_{3}$,
$b_{1}$, $b_{3}$ are time-independent constant. We have 
\begin{equation}
\begin{array}{ccccc}
A_{1}=a_{1},\: & A_{2}=a_{3},\: & B_{1}=b_{1},\: & B_{2}=b_{3},\: & C=c_{1}\end{array}.\label{eq:49}
\end{equation}

In addition, from Eqs.(\ref{eq:25k}, \ref{eq:25m}), and assuming
that there exist $\chi(t)$, $\varphi(t)$, which are time-dependent
matrices, with $\left[\chi(t),\alpha_{2}\right]=\left[\varphi(t),\alpha_{1}\right]=0$.
The time-dependency may appear as follows
\begin{equation}
\begin{array}{c}
b_{1}f_{\Theta}-a_{3}f_{\eta}=\chi(t)\\
b_{3}f_{\Theta}+a_{1}f_{\eta}=\varphi(t)
\end{array}.\label{eq:49-1}
\end{equation}

Now, substituting Eq.(\ref{eq:49}) into Eq.(\ref{eq:25o}) and using
Eq.(\ref{eq:49-1}) give us
\begin{equation}
\frac{\partial c_{1}}{\partial t}=-\left\{ a_{1}f_{\eta}+b_{3}f_{\Theta}\right\} \alpha_{2}-\left\{ b_{1}f_{\Theta}-a_{3}f_{\eta}\right\} \alpha_{1},\label{eq:50}
\end{equation}
using system of relations (\ref{eq:49-1}), we find
\begin{equation}
\frac{\partial c_{1}}{\partial t}=0\mbox{ and }\chi=\varphi=0.\label{eq:51}
\end{equation}
Last but not least, the dynamical invariant (\ref{eq:14}) of time-dependent
NC Dirac equation can be written as follows
\begin{equation}
I=a_{1}p_{x}+b_{1}x+a_{3}p_{y}+b_{3}y+c_{1},\label{eq:52}
\end{equation}
we inferred that Eq.(\ref{eq:13}) is verified and $c_{1}$ should
be a constant. We may note also that all the spin-dependent parts
which are proportional to $\alpha_{j}$, $\beta$ disappear. Which
 means that $I$ has no spin-dependency, but it is proportional to
the matrix of identity in the spinor of space.

\subsection{Eigenvalues and eigenstates of I and $\mathcal{H}(t)$}

Supposing that the invariant in general $I(t)$ is a complete set
of eigenfunctions $\left|\phi(\lambda,k)\right\rangle $ (in this
subsection, the analysis is not concerning only on time-independent
invariants), with $\lambda$ being the corresponding eigenvalue (spectrum
of the operator), and $k$ represents all other necessary quantum
numbers to specify the eigenstates. The eigenvalues equation is written
as
\begin{equation}
I(t)\left|\phi(\lambda,k)\right\rangle =\lambda\left|\phi(\lambda,k)\right\rangle ,\label{eq:53}
\end{equation}
where $\left|\phi(\lambda,k)\right\rangle $ are an orthogonal eigenfunctions
\begin{equation}
\left\langle \phi(\lambda,k)\mid\phi(\lambda^{'},k^{'})\right\rangle =\delta_{\lambda\lambda^{'}}\delta_{kk^{'}}.\label{eq:54}
\end{equation}

According to Eq.(\ref{eq:13-1}), the eigenvalues are real and not
time-dependent. Deriving Eq.(\ref{eq:53}) in time, we find
\begin{equation}
\frac{\partial I}{\partial t}\left|\phi(\lambda,k)\right\rangle +I\frac{\partial}{\partial t}\left|\phi(\lambda,k)\right\rangle =\frac{\partial\lambda}{\partial t}\left|\phi(\lambda,k)\right\rangle +\lambda\frac{\partial}{\partial t}\left|\phi(\lambda,k)\right\rangle ,\label{eq:55}
\end{equation}
we apply Eq.(\ref{eq:13}) over the eigenfunctions $\left|\phi(\lambda,k)\right\rangle $,
we have
\begin{equation}
i\frac{\partial I}{\partial t}\left|\phi(\lambda,k)\right\rangle +I\mathcal{H}^{nc}\left|\phi(\lambda,k)\right\rangle -\mathcal{H}^{nc}\lambda\left|\phi(\lambda,k)\right\rangle =0,\label{eq:56}
\end{equation}
the scalar product of Eq.(\ref{eq:56}) by $\left\langle \phi(\lambda^{'},k^{'})\right|$
is
\begin{equation}
i\left\langle \phi(\lambda^{'},k^{'})\left|\frac{\partial I}{\partial t}\right|\phi(\lambda,k)\right\rangle +\left(\lambda^{'}-\lambda\right)\left\langle \phi(\lambda^{'},k^{'})\left|\mathcal{H}^{nc}\right|\phi(\lambda,k)\right\rangle =0,\label{eq:57}
\end{equation}
which implies
\begin{equation}
\left\langle \phi(\lambda^{'},k^{'})\left|\frac{\partial I}{\partial t}\right|\phi(\lambda,k)\right\rangle =0,\label{eq:58}
\end{equation}
the scalar product of Eq.(\ref{eq:55}) by $\left\langle \phi(\lambda^{'},k^{'})\right|$is
\begin{equation}
\left\langle \phi(\lambda^{'},k^{'})\left|\frac{\partial I}{\partial t}\right|\phi(\lambda,k)\right\rangle =\frac{\partial\lambda}{\partial t},\label{eq:59}
\end{equation}
 from Eq.(\ref{eq:58}), the Eq.(\ref{eq:59}) shows that
\begin{equation}
\left\langle \phi(\lambda^{'},k^{'})\left|\frac{\partial I}{\partial t}\right|\phi(\lambda,k)\right\rangle =\frac{\partial\lambda}{\partial t}=0.\label{eq:60}
\end{equation}

While the eigenvalues are time-independent, the eigenstates should
be time-dependent.

In order to find the link between the eigenstates of the invariant
$I(t)$ and the solutions of the relativistic Dirac equation, firstly,
we start with writing the motion equation of $\left|\phi(\lambda,k)\right\rangle $,
so that using Eq.(\ref{eq:55}) and Eq.(\ref{eq:60}), we obtain
\begin{equation}
\frac{\partial I}{\partial t}\left|\phi(\lambda,k)\right\rangle =\left(\lambda-I\right)\frac{\partial}{\partial t}\left|\phi(\lambda,k)\right\rangle ,\label{eq:61}
\end{equation}
by using the scalar product with $\left\langle \phi(\lambda^{'},k^{'})\right|$,
and taking Eq.(\ref{eq:57}) to eliminate $\left\langle \phi(\lambda^{'},k^{'})\left|\frac{\partial I}{\partial t}\right|\phi(\lambda,k)\right\rangle $,
then we obtain
\begin{equation}
i\left\langle \phi(\lambda^{'},k^{'})\left|\left(\lambda-\lambda^{'}\right)\frac{\partial}{\partial t}\right|\phi(\lambda,k)\right\rangle =\left(\lambda-\lambda^{'}\right)\left\langle \phi(\lambda^{'},k^{'})\left|\mathcal{H}^{nc}\right|\phi(\lambda,k)\right\rangle ,\label{eq:62}
\end{equation}
for $\lambda^{'}\neq\lambda$, we deduce
\begin{equation}
i\left\langle \phi(\lambda^{'},k^{'})\left|\frac{\partial}{\partial t}\right|\phi(\lambda,k)\right\rangle =\left\langle \phi(\lambda^{'},k^{'})\left|\mathcal{H}^{nc}\right|\phi(\lambda,k)\right\rangle ,\label{eq:63}
\end{equation}
then we deduce immediately that $\left|\phi(\lambda,k)\right\rangle $
satisfy the Dirac equation, that is to say $\left|\phi(\lambda,k)\right\rangle $
are particular solutions of Dirac equation.

It is assumed that, a phase has been taken, but it still always possible
to multiply it by an arbitrary time-dependent phase factor, which
means that we can define a new set of $I(t)$ eigenstates linked to
our overall by a time-dependent gauge transformation, and 
\begin{equation}
\left|\phi(\lambda,k)\right\rangle _{\alpha}=e^{i\alpha_{\lambda}(t)}\left|\phi(\lambda,k)\right\rangle ,\label{eq:64}
\end{equation}
where $\alpha_{\lambda}(t)$ is a real time-dependent function arbitrarily
chosen called LR phase, $\left|\phi_{\lambda}(x,y,t)\right\rangle _{\alpha}$are
eigenstates of $I(t)$ which are orthonormal and associated with $\lambda$.
By putting Eq.(\ref{eq:64}) in Eq.(\ref{eq:63}) and using Eq.(\ref{eq:54}),
we find
\begin{equation}
\frac{\partial\alpha_{\lambda,k}}{\partial t}\delta_{\lambda\lambda^{'}}\delta_{kk^{'}}=\left\langle \phi(\lambda^{'},k^{'})\right|i\frac{\partial}{\partial t}-\mathcal{H}^{nc}\left|\phi(\lambda,k)\right\rangle .\label{eq:65}
\end{equation}

All the eigenstates of the invariant are also solutions of the time-dependent
Dirac equation, it was shown in \cite{key-46} that its general solution
is done by
\begin{equation}
\left|\bar{\psi}(t)\right\rangle =\sum_{\lambda,k}C_{\lambda,k}e^{i\alpha_{\lambda,k}(t)}\left|\phi(\lambda,k,t)\right\rangle ,\label{eq:66}
\end{equation}
we remark that Eq.(\ref{eq:66}) is also spin-independent in its state.
But maybe the spin-dependent part is entangled in the coefficient
$C$. $\left|\phi(\lambda,k,t)\right\rangle $ are the orthonormal
eigenstates of $I(t)$, with $C_{\lambda,k}$ are time-independent
coefficients, which correspond to $\left|\psi(0)\right\rangle $
\begin{equation}
C_{\lambda,k}=\left\langle \lambda,k\mid\psi(0)\right\rangle .\label{eq:54-1}
\end{equation}

For a discrete spectrum of $I(t)$, with $\lambda=\lambda^{'}$ ,
$k=k^{'}$, and from Eq.(\ref{eq:65}) the LR phase is defined as
\begin{equation}
\alpha(t)=\int_{0}^{t}\left\langle \phi(\lambda,k,t^{'})\right|i\frac{\partial}{\partial t^{'}}-\mathcal{H}^{nc}(t^{'})\left|\phi(\lambda,k,t^{'})\right\rangle dt^{'}.\label{eq:67}
\end{equation}

But in the continuous spectrum case, the general expression of the
phase is
\begin{equation}
\frac{\partial\alpha_{\lambda,k}}{\partial t}\left\langle \phi(\lambda^{'},k^{'},t^{'}\mid\phi(\lambda,k,t)\right\rangle =\left\langle \phi(\lambda^{'},k^{'},t^{'})\right|i\frac{\partial}{\partial t}-\mathcal{H}^{nc}\left|\phi(\lambda,k,t)\right\rangle ,\label{eq:68}
\end{equation}
where $k$ is an index which varies continuously in the real values,
thus
\begin{equation}
\left\langle \phi(\lambda^{'},k^{'},t^{'}\mid\phi(\lambda,k,t)\right\rangle =\delta_{\lambda\lambda^{'}}\delta(k-k^{'}),\label{eq:68-1}
\end{equation}
substituting Eq.(\ref{eq:68-1}) in Eq.(\ref{eq:68}) yields
\begin{equation}
\alpha(t)=\int\int_{0}^{t}\left\langle \phi(\lambda,k^{'},t^{'})\right|i\frac{\partial}{\partial t^{'}}-\mathcal{H}^{nc}\left|\phi(\lambda,k,t^{'})\right\rangle dt^{'}dk^{'}.\label{eq:69}
\end{equation}

Once found the expression of the phase $\alpha(t)$, we can write
the particular solution of our NC time-dependent Dirac equation (\ref{eq:66}).

We use for simplicity the notation of the discrete spectrum of $I(t)$.
We see that the eigenfunction of $I(t)$ has the form of \cite{key-55,key-57}
\begin{equation}
\left|\phi_{\lambda,k}(x,y,t)\right\rangle \propto\left|\lambda,k\right\rangle exp\left[i\left(\xi_{1}(t)x+\xi_{2}(t)y+\xi_{3}(t)x^{2}+\xi_{4}(t)y^{2}\right)\right],\label{eq:70}
\end{equation}
where $\xi_{1}(t)$, $\xi_{2}(t)$, $\xi_{3}(t)$, $\xi_{4}(t)$ are
arbitrary time-dependent functions.

By substituting Eq.(\ref{eq:70}) into Eq.(\ref{eq:67}) yields
\begin{equation}
\alpha(t)=\vartheta-\int_{0}^{t}E^{nc}dt^{'},\label{eq:71}
\end{equation}
with
\begin{equation}
\vartheta(x,y,t)=\left(\xi_{1}(0)-\xi_{1}(t)\right)x+\left(\xi_{2}(0)-\xi_{2}(t)\right)y+\left(\xi_{3}(0)-\xi_{3}(t)\right)x^{2}+\left(\xi_{4}(0)-\xi_{4}(t)\right)y^{2},\label{eq:71-1}
\end{equation}
and $E^{nc}$ is the eigenvalue of the Hamiltonian (\ref{eq:11}).

Finally, the solution of the NC Dirac equation (\ref{eq:12}) is \cite{key-46}

\begin{equation}
\left|\bar{\psi}(t)\right\rangle =\sum_{\lambda,k}C_{\lambda,k}e^{i[\vartheta-\int_{0}^{t}E^{nc}dt^{'}]}\left|\phi(\lambda,k,t)\right\rangle ,\label{eq:72}
\end{equation}

\subsection{The exact form of the solutions of the problem}

As agreed \cite{key-55,key-56,key-57}, the wave function of the NC
Dirac equation is given by the following trial function 
\begin{equation}
\left|\bar{\psi}(x,y,t)\right\rangle =\mathcal{F}(t)\left|\phi(x,y,t\right\rangle ,\label{eq:73}
\end{equation}
where $\mathcal{F}$ is a time-dependent vector of 2 components ($2\times1$)
\begin{equation}
\mathcal{F}(t)=\left(\begin{array}{c}
\mathcal{F}_{1}(t)\\
\mathcal{F}_{2}(t)
\end{array}\right),\label{eq:74}
\end{equation}
as long as $I(t)$ is independent in time, Eq.(\ref{eq:13-1-1}) goes
to Eq.(\ref{eq:12}). Then the substitution of Eq.(\ref{eq:73}) into
Eq.(\ref{eq:12}), and using Eqs.(\ref{eq:70}, \ref{eq:6-1}) give{\small{}
\begin{equation}
\begin{array}{c}
\left\{ \begin{array}{c}
i\frac{\partial\mathcal{F}_{1}}{\partial t}-\mathcal{F}_{1}\frac{\partial\xi_{1}}{\partial t}x-\mathcal{F}_{1}\frac{\partial\xi_{2}}{\partial t}y-\mathcal{F}_{1}\frac{\partial\xi_{3}}{\partial t}x^{2}-\mathcal{F}_{1}\frac{\partial\xi_{4}}{\partial t}y^{2}\\
i\frac{\partial\mathcal{F}_{2}}{\partial t}-\mathcal{F}_{2}\frac{\partial\xi_{1}}{\partial t}x-\mathcal{F}_{2}\frac{\partial\xi_{2}}{\partial t}y-\mathcal{F}_{2}\frac{\partial\xi_{3}}{\partial t}x^{2}-\mathcal{F}_{2}\frac{\partial\xi_{4}}{\partial t}y^{2}
\end{array}\right\} =\left\{ \begin{array}{cc}
m & \alpha_{1}f_{\Theta}p_{x}-\alpha_{2}f_{\eta}x+\alpha_{2}f_{\Theta}p_{y}+\alpha_{1}f_{\eta}y\\
\alpha_{1}f_{\Theta}p_{x}-\alpha_{2}f_{\eta}x+\alpha_{2}f_{\Theta}p_{y}+\alpha_{1}f_{\eta}y & -m
\end{array}\right\} \\
\times\left(\begin{array}{c}
\mathcal{F}_{1}\\
\mathcal{F}_{2}
\end{array}\right),
\end{array}\label{eq:75}
\end{equation}
}then, we obtain
\begin{equation}
\begin{array}{c}
i\frac{\partial\mathcal{F}_{1}}{\partial t}-\mathcal{F}_{1}\frac{\partial\xi_{1}}{\partial t}x-\mathcal{F}_{1}\frac{\partial\xi_{2}}{\partial t}y-\mathcal{F}_{1}\frac{\partial\xi_{3}}{\partial t}x^{2}-\mathcal{F}_{1}\frac{\partial\xi_{4}}{\partial t}y^{2}=f_{\Theta}\mathcal{F}_{2}p_{x}+if_{\eta}\mathcal{F}_{2}x-if_{\Theta}\mathcal{F}_{2}p_{y}+f_{\eta}\mathcal{F}_{2}y+m\mathcal{F}_{1}\\
i\frac{\partial\mathcal{F}_{2}}{\partial t}-\mathcal{F}_{2}\frac{\partial\xi_{1}}{\partial t}x-\mathcal{F}_{2}\frac{\partial\xi_{2}}{\partial t}y-\mathcal{F}_{2}\frac{\partial\xi_{3}}{\partial t}x^{2}-\mathcal{F}_{2}\frac{\partial\xi_{4}}{\partial t}y^{2}=f_{\Theta}\mathcal{F}_{1}p_{x}-if_{\eta}\mathcal{F}_{1}x+if_{\Theta}\mathcal{F}_{1}p_{y}+f_{\eta}\mathcal{F}_{1}y-m\mathcal{F}_{2}
\end{array},\label{eq:76}
\end{equation}
by solving the above system of equations, we find 
\begin{equation}
\frac{\partial\mathcal{F}_{1}}{\partial t}=-im\mathcal{F}_{1},\quad\frac{\partial\mathcal{F}_{2}}{\partial t}=im\mathcal{F}_{2},\label{eq:77a}
\end{equation}
\begin{equation}
\mathcal{F}_{1}\frac{\partial\xi_{1}}{\partial t}=-i\{\frac{eB}{2}+\frac{\eta}{2}e^{-\gamma t}\}\mathcal{F}_{2},\label{eq:77c}
\end{equation}
\begin{equation}
\mathcal{F}_{1}\frac{\partial\xi_{2}}{\partial t}=-\{\frac{eB}{2}+\frac{\eta}{2}e^{-\gamma t}\}\mathcal{F}_{2},\label{eq:77d}
\end{equation}
\begin{equation}
\frac{\partial\xi_{3}}{\partial t}=\frac{\partial\xi_{4}}{\partial t}=0,\label{eq:77e}
\end{equation}
which lead to obtaining
\begin{equation}
\mathcal{F}_{1}=e^{-imt+q_{1}},\quad\mathcal{F}_{2}=e^{imt+q_{2}},\label{eq:79}
\end{equation}
\begin{equation}
\frac{\partial\xi_{1}}{\partial t}=i\frac{\partial\xi_{2}}{\partial t}=-i\{\frac{eB}{2}+\frac{\eta}{2}e^{-\gamma t}\}e^{i2mt+q_{2}-q_{1}},\label{eq:80}
\end{equation}
\begin{equation}
\xi_{1}=i\xi_{2}=-i\{\frac{\kappa}{4l_{B}^{2}im}e^{i2mt}+\frac{\eta\kappa}{4im-2\gamma}e^{\left(-\gamma+i2m\right)t}\},\label{eq:81}
\end{equation}
with $q_{1}$, $q_{2}$ and $\kappa=e^{q_{2}-q_{1}}$ are real constants,
$l_{B}^{-1}=\sqrt{eB}$ \foreignlanguage{french}{is the magnetic length
\cite{key58}}. In commutative case ($\Theta=\eta=\gamma=0$), then
the above relations (\ref{eq:79}, \ref{eq:81}) return to that of
general quantum mechanics 
\begin{equation}
\xi_{1}(t)=i\xi_{2}(t)=-\frac{\kappa}{4l_{B}^{2}m}e^{i2mt},\quad\left|\phi(x,y,t\right\rangle |_{\eta=\gamma=0}\thicksim e^{-\frac{i\kappa}{4l_{B}^{2}m}e^{i2mt}(x+iy)+o_{1}ix^{2}+o_{2}iy^{2}},\label{eq:82}
\end{equation}
with $o_{1}$, $o_{2}$ are are real constants, and in $t=0$
\begin{equation}
\xi_{1}(t=0)=i\xi_{2}(t=0)=-\frac{\kappa}{4l_{B}^{2}m},\quad and\;\mathcal{F}_{1}=\kappa\mathcal{F}_{2}=e^{q_{1}}.\label{eq:83}
\end{equation}

\section{{\normalsize{}Conclusion }}

In conclusion, the dynamics of the system of time-dependent NC Dirac
equation has been analysed and formulated using LR invariant method.
We introduced the time-dependent noncommutativity using a time-dependent
Bopp-shift translation. Knowing that the NC structure constants postulated
expanding exponentially with the evolution of time, and the time-dependency
have a multitude of other possibilities. 

We benefit from the dynamical invariant following the standard procedure
allowed to construct and to obtain an analytical solution of the system.
Having obtained the explicit solutions could help also to investigate
and reformulate the modified version of Heisenberg\textquoteright s
uncertainty relations emerging from non-vanishing commutation relations
(\ref{eq:1}). The uncertainty for the observables A, B has to satisfy
the inequality {\small{}$\triangle A\triangle B\mid_{\psi}\geq\frac{1}{2}\mid\left\langle \psi\right|\left[A,B\right]\left|\psi\right\rangle \mid$
}with {\small{}$\triangle A\mid_{\psi}^{2}=\left\langle \psi\right|A^{2}\left|\psi\right\rangle -\left\langle \psi\right|A\left|\psi\right\rangle ^{2}$}
and the same for B for any state. Depending on these results, we are
planning to study the pair creation process, and investigate its implications
in quantum optics.
\begin{acknowledgments}
The author wishes to express thanks to Pr Lyazid Chetouani for his
interesting comments and suggestions, also would like to appreciate
anonymous reviewers for their careful reading of the manuscript and
their insightful comments.\end{acknowledgments}


\begin{thebibliography}{10}
\bibitem{key-1}Hartland S. Snyder. Quantized space-time. Phys. Rev.
71 (1947) 38-41. \url{https://doi.org/10.1103/PhysRev.71.38}

\bibitem{key-2}Harland S. Snyder. The Electromagnetic Field in Quantized
Space-Time. Phys. Rev. 72 (1947) 68-71. \url{https://doi.org/10.1103/PhysRev.72.68} 

\bibitem{key-3}A. Connes, M.R. Douglas, and A. Schwarz. Noncommutative
Geometry and Matrix Theory:Compactification on Tori. J. High Energy
Phys. 02 (1998) 003. hep-th/9711162

\bibitem{key-4}T. Banks, W. Fischler, S. Shenker and L. Susskind.
M theory as a matrix model: A conjecture. Phys. Rev. D55, 5112 (1997).
\url{https://doi.org/10.1103/PhysRevD.55.5112}

\bibitem{key-5}M. Chaichian, M.M. Sheikh-Jabbari, and A. Tureanu.
Hydrogen Atom Spectrum and the Lamb Shift in Noncommutative QED. Phys.
Rev. Lett. 86 (2001) 2716. \url{https://doi.org/10.1103/PhysRevLett.86.2716}

\bibitem{key-6}B. Mirza and M. Zarei. Non-commutative quantum mechanics
and the Aharonov-Casher effect. Eur. Phys. J. C32(2004) 583. \url{https://doi.org/10.1140/epjc/s2003-01522-8}

\bibitem{key-7}C. Duval and P.A. Horvathy. The exotic Galilei group
and the \textquotedblleft Peierls substitution\textquotedblright .
Phys. Lett. B. 479 (2000) 284. \url{https://doi.org/10.1016/S0370-2693(00)00341-5}

\bibitem{key-8}A. Connes. Non-commutative differential geometry.
Publications Mathématiques de l'IHÉS. 62 (1985) 41. \url{https://doi.org/10.1007/BF02698807}

\bibitem{key-9}A. Connes. A short survey of noncommutative geometry.
J. Math. Phys. 2000, 41, 3832. \url{https://doi.org/10.1063/1.533329}

\bibitem{key-10}S.L. Woronowicz. Twisted SU(2) Group. An Example
of a Non-Commutative Differential Calculus. Publ. Res. Inst. Math.
Sci. 23 (1987) 117. \url{https://doi.org/10.2977/prims/1195176848}

\bibitem{key-11}Seiberg, N.; Witten, E. String theory and noncommutative
geometry. J. High Energy Phys. 1999, 9. \url{https://doi.org/10.1088/1126-6708/1999/09/032}

\bibitem{key-12}M. J., Schraml, S., Schupp, P. et al. Eur. Phys.
J. C. 16 (2000) 161. \url{https://doi.org/10.1007/s100520050012}

\bibitem{key-13}Haouam, I. (2019) Continuity Equation in Presence
of a Non-Local Potential in Non-Commutative Phase-Space. Open Journal
of Microphysics, 9, 15-28. \url{https://doi.org/10.4236/ojm.2019.93003}

\bibitem{key-14}Haouam, I. On the Fisk\textendash Tait equation for
spin-3/2 fermions interacting with an external magnetic field in noncommutative
space-time. Journal of Physical Studies 2020, 24, 1801. \url{https://doi.org/10.30970/jps.24.1801}

\bibitem{key-15}Haouam, I. The Non-Relativistic Limit of the DKP
Equation in Non-Commutative Phase-Space. Symmetry 2019, 11, 223. \url{https://doi.org/10.3390/sym11020223} 

\bibitem{key-16}Haouam, I. and Chetouani, L. (2018) The Foldy-Wouthuysen
Transformation of the Dirac Equation in Noncommutative Phase-Space.
Journal of Modern Physics, 9, 2021-2034. \url{https://doi.org/10.4236/jmp.2018.911127}

\bibitem{key-17}J. B., Möller, L., Schraml, S. et al. Construction
of non-Abelian gauge theories on noncommutative spaces. Eur. Phys.
J. C. 21 (2001) 383. \url{https://doi.org/10.1007/s100520100731}

\bibitem{key-18}Kamoshita, J. Probing noncommutative space-time in
the laboratory frame. Eur. Phys. J. C (2007) 52: 451. \url{https://doi.org/10.1140/epjc/s10052-007-0371-y}

\bibitem{key-19}Alisultanov Z. Z. Landau levels in graphene in crossed
magnetic and electric fields: Quasi-classical approach. Physica B,438
(2014) 41. \url{https://doi.org/10.1016/j.physb.2013.12.033}

\bibitem{key-20}Alisultanov Z. Z. Oscillations of magnetization in
graphene in crossed magnetic and electric fields. JETP Lett. 99 (2014)
232. \url{https://doi.org/10.1134/S0021364014040055}

\bibitem{key-21}Ma N., Zhang S., Liu D.and Wang V. Influence of electrostatic
field on the Weiss oscillations in graphene. Phys. Lett. A,378 (2014)
3354. \url{https://doi.org/10.1016/j.physleta.2014.09.026}

\bibitem{key-22}Zhang S., Ma N.andZhang E.,J. Phys.: Condens.Matter,22
(2010) 115302.

\bibitem{key-23}Peres N. M. R.andCastro E. V.,J. Phys.: Condens.Matter,19
(2007) 406231.

\bibitem{key-24}Novoselov K. S., Geim A. K., Morozov S. V., JiangD.,
Katsnelson M. I., Grigorieva I. V., Dubonos S.V.andFirsov A. A.,Nature,438
(2005) 197

\bibitem{key-25}De Martino A., Dell\textquoteright Anna L.andEgger
R.,Phys.Rev. Lett.,98 (2007) 066802.

\bibitem{key-26}Dell\textquoteright Anna L.andDe Martino A.,Phys.
Rev B,79 (2009) 045420.

\bibitem{key-27}Zhang Y., Tan Y.-W., Stormer H. L.andKim P.,Nature,438
(2005) 201.

\bibitem{key-28}Bolotin K. I., Ghahari F., Shulman M. D., StormerH.
L.andKim P.,Nature,462 (2009) 196.

\bibitem{key-29}Mikitika G. P.andSharlaib Yu. V.,Low Temp. Phys.,34
(2008) 794.

\bibitem{key-30}Lukose V., Shankar R.andBaskaran G.,Phys. Rev.Lett.,98
(2007) 116802.

\bibitem{key-31}Vicente A, Eduardo V. Castro, and María A. H. Vozmediano
Phys. Rev. B 96 (2017), 081110 (R). \url{https://doi.org/10.1103/PhysRevB.96.081110}

\bibitem{key-32}G. Burmeister and K. Maschke. Scattering by time-periodic
potentials in one dimension and its influence on electronic transport.
Phys. Rev. B 57 (1998) 13050. \url{https://doi.org/10.1103/PhysRevB.57.13050 }

\bibitem{key-33}C. S. Tang and C. S. Chu. Coherent quantum transport
in narrow constrictions in the presence of a finite-range longitudinally
polarized time-dependent field. Phys. Rev. B 60 (1999) 1830. \url{https://doi.org/10.1103/PhysRevB.60.1830}

\bibitem{key-34}W. Li and L. E. Reichl. Transport in strongly driven
heterostructures and bound-state-induced dynamic resonances. Phys.
Rev. B 62 (2000) 8269. \url{https://doi.org/10.1103/PhysRevB.62.8269}

\bibitem{key-35}C. Figueira de Morisson Faria, M. Dörr, and W. Sandner.
Time profile of harmonic generation. Phys. Rev. A 55, 3961 (1997).
\url{https://doi.org/10.1103/PhysRevA.55.3961}

\bibitem{key-36}H. Zeng. Quantum-state control in optical lattices.
Phys. Rev. A 57, 388 (1997). \url{https://doi.org/10.1103/PhysRevA.57.1972}

\bibitem{key-37}H. Maeda and T. F. Gallagher. Nondispersing Wave
Packets. Phys. Rev. Lett. 92 (2004) 133004. \url{https://doi.org/10.1103/PhysRevLett.92.133004}

\bibitem{key-38}C. E. Creffield and G. Platero. ac-driven localization
in a two-electron quantum dot molecule. Phys. Rev. B 65 (2002) 113304.
\url{https://doi.org/10.1103/PhysRevB.65.113304} 

\bibitem{key-39}Horace P. Yuen. Two-photon coherent states of the
radiation field. Phys. Rev. A 13 (1976) 2226. \url{https://doi.org/10.1103/PhysRevA.13.2226}

\bibitem{key-40}M. Governale, F. Taddei, and R. Fazio. Pumping spin
with electrical fields. Phys. Rev. B 68, 155324 (2003). \url{https://doi.org/10.1103/PhysRevB.68.155324} 

\bibitem{key-41}A. G. Mal\textquoteright shukov, C. S. Tang, C. S.
Chu, and K. A. Chao. Spin-current generation and detection in the
presence of an ac gate. Phys. Rev. B 68, 233307 (2003). \url{https://doi.org/10.1103/PhysRevB.68.233307}

\bibitem{key-42}L.S.Brown. Quantum motion in a Paul trap. Phys.Rev.Lett.,
66, 527(1991). \url{https://doi.org/10.1103/PhysRevLett.66.527}

\bibitem{key-43}Mang Feng Phys. Rev. A 64, 034101 Complete solution
of the Schrödinger equation for the time-dependent linear potential.
\url{https://doi.org/10.1103/PhysRevA.64.034101}

\bibitem{key-44}Liang, ML., Zhang, ZG. \& Zhong, KS. Czech J Phys
(2004) 54: 397. \url{https://doi.org/10.1023/B:CJOP.0000020579.42018.d9}

\bibitem{key-45}Lewis H R Jr. Classical and Quantum Systems with
Time-Dependent Harmonic-Oscillator-Type Hamiltonians. Phys. Rev. Lett.
18 510 (1967). \url{https://doi.org/10.1103/PhysRevLett.18.510}

\bibitem{key-46}Lewis H R Jr. and Riesenfeld W B. An Exact Quantum
Theory of the Time-Dependent Harmonic Oscillator and of a Charged
Particle in a Time-Dependent Electromagnetic Field. J. Math. Phys.
10 1458 (1969). \url{ https://doi.org/10.1063/1.1664991} 

\bibitem{key-47}Pedrosa I A, Melo J L and Nogueira E Jr. Linear invariants
and the quantum dynamics of a nonstationary mesoscopic RLC circuit
with a source. Mod. Phys. Rev. Lett. B28 1450212 (2014). \url{https://doi.org/10.1142/S0217984914502121}

\bibitem{key-48}Chen X, Ruschhaupt A, Schmidt S, del Campo A, Guéry
Odelin D and Muga JG. Fast Optimal Frictionless Atom Cooling in Harmonic
Traps: Shortcut to Adiabaticity. Phys. Rev. Lett. 104 063002 (2010).
\url{https://doi.org/10.1103/PhysRevLett.104.063002}

\bibitem{key-49}\foreignlanguage{french}{Dey, Sanjib et al. Noncommutative
quantum mechanics in a time-dependent background. Phys.Rev. D90 (2014)
no.8, 084005. \url{https://doi.org/10.1103/PhysRevD.90.084005}}

\bibitem{key-50}V. V. Nesvizhesky et al. Measurement of quantum states
of neutrons in the Earth\textquoteright s gravitational field. Phys.
Rev. D67, 102002 (2003). \url{https://doi.org/10.1103/PhysRevD.67.102002}

\bibitem{key-51}V. V. Nesvizheskyet al. Quantum states of neutrons
in the Earth's gravitational field. Nature (London) 415, 297 (2002). 

\bibitem{key-52}S. M. Carroll, J. A. Harvey, V. A. Kostelecký, C.
D. Lane, and T. Okamoto. Noncommutative Field Theory and Lorentz Violation.
Phys. Rev. Lett.87, 141601(2001). \url{https://doi.org/10.1103/PhysRevLett.87.141601}

\bibitem{key-53}O. Bertolami, J. G. Rosa, C. M. L. de Aragão, P.
Castorina, and D. Zappalà. Noncommutative gravitational quantum well.
Phys. Rev. D 72, 025010 (2005). \url{https://doi.org/10.1103/PhysRevD.72.025010}

\bibitem{key-54}\foreignlanguage{french}{F Delduc et al 2008 J. Phys.:
Conf. Ser. 103 012020. \url{https://doi.org/10.1088/1742-6596/103/1/012020}}

\bibitem{key-55}X. Jiang, C. Long and S. Qin. Solution of Dirac Equation
with the Time-Dependent Linear Potential in Non-Commutative Phase
Space. J. Mod. Phys. 4 (2013) 940. \url{https://doi.org/10.4236/jmp.2013.47126}

\bibitem{key-56}M. Merad, S. Bensaida. Wave functions for a Duffin-Kemmer-Petiau
particle in a time-dependent potential. J. Math. Phy. 48, 073515 (2007).
\url{https://doi.org/10.1063/1.2747609}

\bibitem{key-57}H. Sobhani and H. Hassanabadi. Two-Dimensional Linear
Dependencies on the Coordinate Time-Dependent Interaction in Relativistic
Non-Commutative Phase Space. Commun. Theor. Phys. 64 (2015) 263. \url{https://doi.org/10.1088/0253-6102/64/3/263}

\bibitem{key58}\foreignlanguage{french}{Z. Jiang, E. A. Henriksen,
L. C. Tung, Y.-Y. Wang, M. E. Scharwtz, M. Y. Han, P. Kim and H. L.
Stormer, Phys. Rev. Lett. 98, 197403, (2007). \url{https://doi.org/10.1103/PhysRevLett.98.197403}}\end{thebibliography}
\end{document}